\documentclass[preprintnumbers, floatfix,letterpaper,aps,prd,epsfig,nofootinbib,twocolumn]{revtex4-1}
\pdfoutput=1
\usepackage{bm,graphicx,dcolumn,epstopdf,epsf, latexsym,mathbbol, amssymb,amsmath,color,slashed, mathrsfs,mathcomp, simplewick}
\usepackage[center]{subfigure}
\usepackage{multirow}
\usepackage{makecell}
\usepackage{color, soul}
\usepackage[colorlinks,linkcolor=blue,citecolor=blue,urlcolor=blue]{hyperref}
\sethlcolor{yellow}
\setstcolor{blue}
\setulcolor{red}

\begin{document}
	\allowdisplaybreaks
	\newcommand{\bq}{\begin{equation}}
	\newcommand{\eq}{\end{equation}}
	\newcommand{\bqn}{\begin{eqnarray}}
	\newcommand{\eqn}{\end{eqnarray}}
	\newcommand{\nb}{\nonumber}
	\newcommand{\lb}{\label}
	\newcommand{\f}{\frac}
	\newcommand{\p}{\partial}
	\newcommand{\PRL}{Phys. Rev. Lett.}
	\newcommand{\PLB}{Phys. Lett. B}
	\newcommand{\PRD}{Phys. Rev. D}
	\newcommand{\CQG}{Class. Quantum Grav.}
	\newcommand{\JCAP}{J. Cosmol. Astropart. Phys}
	\newcommand{\JHEP}{J. High. Energy. Phys.}
	\newcommand{\red}{\textcolor{black}}

	\title{Constraints on self-dual black hole in loop quantum gravity with S0-2 star in the Galactic Center}
	
	\author{Jian-Ming Yan${}^{a, b}$}
	\email{yanjm@zjut.edu.cn}

	\author{Qiang Wu${}^{a, b}$}
	\email{wuq@zjut.edu.cn}
   
   \author{Cheng Liu${}^{a,b}$}
   \email{liucheng@zjut.edu.cn}
	
	\author{Tao Zhu${}^{a, b}$}
	\email{zhut05@zjut.edu.cn; Corresponding author}
	
	\author{Anzhong Wang${}^{c}$}
	\email{anzhong\_wang@baylor.edu}
	
	\affiliation{${}^{a}$Institute for Theoretical Physics \& Cosmology, Zhejiang University of Technology, Hangzhou, 310023, China\\
		${}^{b}$ United Center for Gravitational Wave Physics (UCGWP),  Zhejiang University of Technology, Hangzhou, 310023, China\\
		${}^{c}$ GCAP-CASPER, Physics Department, Baylor University, Waco, Texas 76798-7316, USA}
	
	\date{\today}
	
\begin{abstract}
One of remarkable features of loop quantum gravity (LQG) is that it can provide resolutions to both the black hole and big bang singularities. In the mini-superspace approach based on the polymerization procedure in LQG, a \red{quantum corrected black hole metric} is constructed. \red{This metric is also known as self-dual spacetime since the form of the metric is invariant under the exchange $r \to a_0/r$ with $a_0$ being proportional to the minimum area in LQG and $r$ is the standard radial coordinate at asymptotic infinity.} It modifies the Schwarzschild spacetime by the polymeric function $P$, purely due to the geometric quantum effects from LQG. \red{Here $P$ is related to the polymeric parameter $\delta$ which is introduced to define the paths one integrates the connection along to define the holonomies in the quantum corrected Hamiltonian constraint in the polymerization procedure in LQG.} In this paper, we consider its effects on the orbital signatures of S0-2 star orbiting Sgr A* in the central region of our Milky Way, and compare it with the publicly available astrometric and spectroscopic data, including the astrometric positions, the radial velocities, and the orbital \red{precession} for the S0-2 star. We perform Monte Carlo Markov Chain (MCMC) simulations to probe the possible LQG effects on the orbit of S0-2 star. \red{No significant evidence of the self-dual spacetime arisIng from LQG is found. We thus place an upper bounds at 95\% confidence level on the polymeric function $P < 0.043$ and $P < 0.056$, for Gaussian and uniform priors on orbital parameters, respectively. }

\end{abstract}

\maketitle
	
\section{Introduction}
\renewcommand{\theequation}{1.\arabic{equation}} \setcounter{equation}{0}

Although general relativity (GR) \red{is} considered to be the most successful theory of gravity since it was proposed, it faces difficulties both theoretically (e.g. singularity, quantization, etc), and observationally (e.g. dark matter, dark energy, etc). In particular, Einstein’s GR does not employ any quantum principles and it is still an unsolved question of unifying GR and quantum mechanics \cite{QG1, QG2}. GR also inevitably leads to singularities both at the \red{beginning} of the universe \cite{singularity1, singularity2} and in the interiors of black hole spacetimes \cite{hawking} at which our known physics laws become all invalid.  Various modified gravities or quantum gravities have been proposed to be one of the effective ways to solve these anomalies. Therefore, the tests of the modified gravities beyond GR are essential to \red{ constrain alternative theories of  gravity.}

LQG provides \red{remarkable} resolution of both the classical big bang and black hole singularities. \red{In loop quantum cosmology (LQC), the big bang singularity is replaced by a quantum bounce and the universe that starts at the bounce can eventually evolve to the current stage of the universe \cite{LQC1, LQC2}. In the interior of black hole, the singularity can be solved due to the existence of a minimal area gap in LQG, see \cite{LQG_BH, Ashtekar:2005qt} for example.} Recently, a regular static spacetime metric, the self-dual spacetime, is derived in mini-superspace approach based on the polymerization procedure in LQG \cite{LQG_BH}. This self-dual spacetime is regular and free of any spacetime curvature singularity. In this spacetime, the effects of LQG are characterized by two parameters, the minimal area and the Barbero-Immirzi parameter, arising from LQG. \red{One can verify that under the transformation $ r \to a_0/r$ with $a_0$ being related to the minimal area gap of LQG, the metric remains invariant, with suitable re-parameterization of other variables, hence marking itself as satisfying the T-duality \cite{Sahu:2015dea, Modesto:2009ve}.} This is also the reason that we call it the self-dual spacetime. In the last couple of years, black holes in LQG have been extensively studied, see, for instance, \cite{AOS18a, AOS18b, BBy18, ABP19, ADL20, Gambini:2020qhx, Garcia-Quismondo:2021xdc} and references therein.  For more details, we refer the reader to the review articles, \cite{Perez17, Rovelli18, BMM18, Ashtekar20, Gan:2020dkb}.

 It is natural to ask  whether the LQG effects on the self-dual spacetime can leave any observational signatures for the current and/or forthcoming experiments, so LQG can be tested or constrained directly. Such considerations have resulted in a flourish of studies during the past decades from different kinds of experiments and observations \cite{Alesci:2011wn, Chen:2011zzi, Dasgupta:2012nk, Hossenfelder:2012tc, Barrau:2014yka, Sahu:2015dea, Cruz:2015bcj, add1, add2, Cruz:2020emz, Santos:2021wsw, Liu:2020ola, Sahu:2015dea, Zhu:2020tcf}. In \cite{Liu:2020ola},  the LQG effects on the shadow of the rotating black hole has been discussed in details and their observational implications in comparing with the latest Event Horizon Telescope (EHT) observation of the supermassive black hole, M87* has also been explored \cite{Liu:2020ola}. In addition, the gravitational lensing in the self-dual spacetime has also been studied and the polymeric function from LQG has been constrained by using the Geodetic Very-Long-Baseline Interferometry Data of the solar gravitational deflection of Radio Waves \cite{Sahu:2015dea}. Recently, the solar system test of the self-dual spacetime have  been consideried in \cite{Zhu:2020tcf}, from which the observational constraints on the polymeric function $P$ of LQG are derived as well. It is interesting to note that the phenomenological study of other types of loop quantum black holes/ quantum black holes have also been extensively explored, see \cite{Liu:2021djf, Daghigh:2020fmw, Bouhmadi-Lopez:2020oia, Fu:2021fxn, Brahma:2020eos, Giddings:2019jwy, Giddings:2016btb, Giddings:2017jts, Barrau:2019swg, Haggard:2016ibp, del-Corral:2022kbk} and references therein.

On the other hand, there \red{is strong evidence} that a supermassive black hole inhabits the center of our own Milky Way galaxy \cite{sgA1, sgA2}. It is coincident with a very compact and variable X-ray, infrared, and radio source, Sgr A*, which in turn is surrounded by a very dense cluster of orbiting young and old stars \cite{sgA2, sgA3}. Stars orbiting around Sgr A* have been detected and monitored through the last three decades \cite{sgA2, sgA3}. They move with large velocities (1000 km/s) in Keplerian orbits, pointing out that in the centre of the Galaxy must reside a compact object of mass of $M \sim 4 \times 10^6 M_\odot$  concentrated within a few hundreds Schwarzschild radii. Here $M_{\odot}$ denotes the solar mass. The monitoring of the stellar cluster orbiting around Sgr A*  provides us a great opportunity to test the predictions of general relativity in the regime of strong gravity and improve our understanding about the geometrical properties of the supermassive black hole. 


The gravitational redshift from Sgr A* has been detected at high significance in the spectrum of the star S0-2 during the 2018 pericenter passage of its 16-year orbit \cite{GRAVITY1, GRAVITY1a, DoSci}. The S0-2 star is a B-type star in the nuclear cluster orbiting the radio source Sgr A* in the galactic center with a orbit that is characterized by an orbital period of 16 years, a semi-major axis of 970 AU and an high eccentricity of 0.88 \cite{GRAVITY1, GRAVITY1a, DoSci}. Recently, the relativistic Schwarzschild precession of the pericenter has also been detected in S0-2's orbit \cite{GRAVITY2}. These results are in good agreement with the predictions of general relativity by assuming the geometry of black hole is described by the Schwarzschild metric and  \red{no  significant deviation} from GR has been found. Importantly, these precise observations can also provide a significant way to probe the matter environment surrounding the black hole and distinguish or constrain black holes in different gravity theories. Along this direction, a lot of works have been already carried out. For instance, testing the no-hair theorem with Srg A* \cite{nohair, nohair2}, the studies of a black hole with dark energy interaction \cite{Benisty:2021cmq} and surrounded by dark matter \cite{Zakharov:2021cgx, scalarDM, Heissel:2021pcw, dmspike,Becerra-Vergara:2021gmx}, testing GR \cite{Fang:2020pjm, testGR}, fitting of the orbital motion of S0-2 star in different theories \cite{Borka:2021omc, deMartino:2021daj, DellaMonica:2021xcf, DellaMonica:2021fdr, DAddio:2021smm}, etc.

In this paper, we consider the effects of the self-dual spacetime of LQG on the orbit of S0-2 star orbiting Sgr A* in the central region of our Milky Way. The effects of LQG may not only lead to signatures on the orbits of S0-2 star, but also affect its overall pericentre advance. We also compare the orbit of the self-dual spacetime with the publicly available astrometric and spectroscopic data, including the astrometric positions, the radial velocities, and the orbital \red{precession} for the S0-2 star. With these data, we perform a MCMC simulation to probe the possible LQG effects on the orbit of S0-2 star. \red{We consider two different priors for all the 13 orbital parameters of S0-2 star, the Gaussian and the uniform prior, respectively. For the LQG parameters, we choose uniform priors for both simulations. From these simulations, we did not found any significant evidence of LQG effects and thus placed upper bounds at 95\% confidence level on the polymeric function $P < 0.043$ and $P<0.056$, for Gaussian and uniform priors on orbital parameters, respectively. 
These bounds lead to  constraints on the polymeric parameter $\delta$ of LQG to be $\delta< 1.82$ and $\delta< 2.11$ respectively.} 
At last, we would like to mention that we only consider the static self-dual spacetime in this paper and ignore the effects of the angular momentum of the spacetime. For all the observational effects we considered in this paper, the effects due to rotation of the central black hole are expected to be very small.
 
The plan of our paper is as follows. In Sec. II, we present a very brief introduction of the self-dual spacetime and the \red{geodesic} motion of a massive test particle in this spacetime. With this spacetime metric, in Sec. III, we first briefly summarize the publicly available astrometric and spectroscopic data used in this paper, including the astrometric positions, the radial velocities, and the orbital \red{precession} for the S0-2 star. And then we build the orbital model and contrast it to the data \red{used in }a MCMC simulation. We discuss the main results of our analysis in Sec. IV. A summary of our main conclusions of this paper is presented in Sec. V.

\section{Equation of motion  for test particles in the self-dual spacetime \label{secrot}}
\renewcommand{\theequation}{2.\arabic{equation}} \setcounter{equation}{0}
	
\subsection{Self-dual spacetime}

\red{In this subsection, we provide a brief introduction of the self-dual spacetime in LQG  proposed in \cite{LQG_BH}. This spacetime is a quantum corrected Schwarzschild spacetime, and the effects of LQG are characterized by two parameters, the minimal area and the polymeric parameter $\delta$, arising from LQG. In order to study LQG effects in the Schwarzschild spacetime, one starts with the Kantowski-Sachs spacetime, }
\red{
\bqn
ds^2 = - N^2(t) d t^2 + \frac{p_b^2}{|p_c| L_0^2}dx^2 + |p_c|d^2\Omega,\lb{KS}
\eqn
where $d^2\Omega = d\theta^2 + \sin^2\theta d\phi^2$, and $L_0$ is the length of the fiducial cell with $c \in (0, L_0)$. The quantities $b$, $c$, $p_b$, and $p_c$ represent the dynamical variables of the spacetime. The Kantowski-Sachs spacetime is isometric to the interior of the Schwarzschild spacetime \cite{AOS18a}. Thus, if one uses Hamiltonian constraint ${\cal C}_{H}$ of general relativity (GR),  the classical Schwarzschild spacetime inside the event horizon can be produced from the dynamical trajectories on phase space \cite{LQG_BH}.}

\red{The Kantowski-Sachs spacetime, which is given in (\ref{KS}) and isometric to the interior of the Schwarzschild spacetime, can also be used to describe a homogeneous but anisotropic universe. It is this reason that one can directly apply the similar quantization procedure from loop quantum cosmology (LQC) to deal with the singularity in the interior of the Schwarzschild spacetime. In the treatment of LQC, the full quantum evolution is extremely well approximated by certain quantum corrected effective equations. Similar treatment is applied to the interior of the Schwarzschild spacetime to get the quantum corrected Schwarzschild spacetime which cues the black hole singularity, see \cite{LQG_BH, Ashtekar:2005qt, AOS18a} and references therein. }


\red{There are two key ingredients in the quantization procedure of black hole in LQG. One is the existence of the minimal area $A_{\rm min}=4\sqrt{3} \pi \gamma l_{\rm Pl}^2$ with $\gamma$ being to the Immirzi parameter and $l_{\rm Pl}$ the Planck length, which represents the minimum non-zero eigenvalue of the area operator.  Another ingredient is the mini-superspace polymer-like quantization inspired by LQG, in which the effective quantum theory is achieved by replacing the canonical variables $(b, c)$ in the phase space with their regularized ones \cite{xxx},
\bqn
b \to \frac{\sin(\delta_b b)}{\delta_b},\;\;\; c\to \frac{\sin(\delta_c c)}{\delta_c},
\eqn
where $\delta_b$ and $\delta_c$ are two different polymeric parameters introduced to define the lengths of the paths along which we integrate to define the holonomies. The two parameters control at which scales the quantum effect is relevant. It is easy to see that when $\delta_b,\; \delta_c \to 0$, the classical limit is recovered. However, due to the lack of the complete theory of quantum gravity, a full picture on how to chose $\delta_b$ and $\delta_c$ is still absent. In the literature, there are a lot of different choices from different perspectives (see \cite{Gan:2020dkb} and references therein). }

\red{In this paper we consider a quantum corrected black hole from the possible choice of treating $\delta_b$ and $\delta_c$ as constants. This choice is also called $\mu_0$-scheme in LQG and has been adopted in deriving the effective LQG black hole in \cite{LQG_BH, Ashtekar:2005qt, Modesto:2005zm, Campiglia:2007pb, ADL20}. With such choices of $\delta_b$ and $\delta_c$, from the effective quantum Hamiltonian constraint one can solve the equation of motion to get an effective quantum corrected Schwarzschild black hole, as performed in \cite{LQG_BH} for example. The effective solution obtained in this way is only valid in the interior of the quantum corrected Schwarzschild spacetime. As pointed out in \cite{LQG_BH, Modesto:2009ve},  an analytic continuation to the region outside the horizon shows that one can reduce the two free parameters by identifying the minimum area in the solution with the minimum area of LQG. The remaining unknown constant of the model, $\delta_b$, is the dimensionless polymeric parameter and must be constrained by experiment\cite{LQG_BH, Modesto:2009ve}. It is the main purpose of this paper to derive observational constraint by using the astrometric data of S0-2 star in the Galactic center.}

\red{The quantum corrected Schwarzschild spacetime derived from the above procedure} has also been shown to be geodesically complete and free of any spacetime curvature singularity. The metric of this spacetime is given by \cite{LQG_BH}
	\begin{equation}\label{1}
	ds^2= - f(r)dt^2 + \frac{dr^2}{g(r)} + h(r)(d\theta^2+\sin^2\theta d\phi^2),
	\end{equation}
where the metric functions $f(r)$, $g(r)$, and $h(r)$ are given by
	\begin{eqnarray}
	f(r)&=&\frac{(r-r_+)(r-r_-)(r+r_*)^2}{r^4+a_0^2},\nonumber\\
	g(r)&=&\frac{(r-r_+)(r-r_-)r^4}{(r+r_*)^2(r^4+a_0^2)},\nonumber\\
	h(r)&=&r^2+\frac{a_0^2}{r^2}. \lb{metric}
	\end{eqnarray}
Here $r_+=2 G M/(1+P)^2$ and $r_{-} = 2G M P^2/(1+P)^2$ are the two horizons, and $r_{*}= \sqrt{r_+ r_-} = 2G MP/(1+P)^2$ with $G$ representing the gravitational constant, $M$ denoting the ADM mass of the solution, and $P$ being the polymeric function \red{ of the polymeric parameter $ \delta$ \footnote{Hereafter we use $\delta=\delta_b$ since now there is only one independent polymeric parameter.}.}
\bqn
P \equiv \frac{\sqrt{1+\epsilon^2}-1}{\sqrt{1+\epsilon^2}+1},
\eqn
where $\epsilon$ denotes a product of the Immirzi parameter $\gamma$ and the polymeric parameter $\delta$ which is introduced to define the paths one integrates the connection along to define the holonomies in the quantum corrected Hamiltonian constraint in the polymerization procedure in LQG \cite{LQG_BH, Modesto:2009ve}. As mentioned in \cite{add1}, the parameter $\epsilon$ (or equivalently the polymeric function $P$) is in principle unbounded but the the procedure for \red{getting} the effective metric is rigorous only when $\epsilon=\gamma \delta \ll 1$. 

\red{Here we want to explain why this metric is self-dual. One can do a transformation $r \rightarrow R=\frac{a_0}{r}, r_{-} \rightarrow R_{-}=\frac{a_0}{r_{-}}, r_{+} \rightarrow R_{+}=\frac{a_0}{r_{+}}  $, with a a rescaling of the time coordinate $t \rightarrow (a_{0}r_{+}^{1/2}r_{-}^{1/2}r_{*})t$. Then we can get 
\begin{eqnarray}
	f(R)&=&\frac{(R-R_+)(R-R_-)(R+R_*)^2}{R^4+a_0^2},\nonumber\\
	g(R)&=&\frac{(R-R_+)(R-R_-)R^4}{(R+R_*)^2(R^4+a_0^2)},\nonumber\\
	h(R)&=&R^2+\frac{a_0^2}{R^2}.
\end{eqnarray}
Thus the metric remains invariant under the transformtion, hence marking itself as satisfying the T-duality \cite{Sahu:2015dea, Modesto:2009ve}.}

The parameter $a_{0}$ in the above metric is defined as
\bqn
a_0 = \frac{A_{\rm min}}{8\pi},
\eqn
where $A_{\rm min}$ represents the minimum area gap of LQG. It is interesting to mention that $A_{\rm min}$ is related to the Planck length $l_{\rm Pl}$ through $A_{\min} \simeq 4 \pi \gamma \sqrt{3} l_{\rm Pl}^2$ \cite{Modesto:2009ve,Sahu:2015dea}. Thus, $a_0$ is proportional to $l_{\rm Pl}$ and is expected to be negligible. Hence, phenomenologically,  the effects of $a_0$ on the spacetime are expected to be very small at the scale of the solar system, and in this paper we will only focus on the solar system effects of the polymeric function $P$ and set $a_0=0$.

Another parameter, the Immirzi parameter $\gamma$, its value has a lot of choices from different considerations, see \cite{Achour:2014rja, Frodden:2012dq, Achour:2014eqa, Han:2014xna, Carlip:2014bfa, Taveras:2008yf} and references therein. It has been shown that its value can even be complex \cite{Achour:2014rja, Frodden:2012dq, Achour:2014eqa, Han:2014xna, Carlip:2014bfa}, or considered as a scalar field in which the value would be fixed by the dynamics \cite{Taveras:2008yf}. In this paper, in order to derive the observational constraints on the polymeric parameter $\delta$ derive from the constraints on $P$, we adopt the commonly used value $\gamma = 0.2375$  from the black hole entropy calculation \cite{immi}.  In order to recover Newtonian limit, one can relate the parameter $G$ with the Newtonian's gravitational constant $G_{\rm N}$ as $G_{N}=G(1+4P)$. For later convenience, we set $G_N=1$ hereafter. 

Here we also \red{need} to mention that when all the effects of LQG is absent, the metric of the self-dual spacetime reduces to the Schwarzschild spacetime exactly.  
	
\subsection{Geodesic equation and orbital \red{precession} of a massive test particle in the self-dual spacetime}

Our purpose here is to study the motion of massive test particles which \red{follow} the time-like geodesics in the self-dual spacetime. The time-like geodesic equations of the metric (\ref{metric}) reads
\bqn
\frac{d^2x^\mu}{d\lambda^2} + \Gamma^\mu_{\nu \rho} \frac{dx^\nu}{d\lambda} \frac{dx^\rho}{d\lambda}=0,\lb{geodesic}
\eqn
where $\lambda$ denotes the affine parameter of the world line of the particle and $\Gamma^{\mu}_{\nu \rho}$ are the Christoffel symbols of the self-dual spacetime. \red{For massive particles, $\lambda$ should be proper time $\tau$.}
For spherical symmetric spacetime, the \red{motions} of the massive particles are confined to a plane so we can set the orbital plane as the equatorial plane and $\theta=\pi/2$ without loss of generality. 

\red{Since the spacetime we considered is static and spherical symmetric, it has two Killing vectors, $\xi_{(t)}^{\mu} = \frac{\partial x^\mu}{\partial t} = (1, 0, 0, 0)$ and $\xi_{\phi}^{\mu} = \frac{\partial x^\mu}{\partial \phi} = (0,0,0,1)$. These two Killing vectors correspond to two conserved quantities, the energy per unit mass $\tilde E$ and the angular momentum per unit mass $\tilde l $ of the massive particle. With these two Killing vectors, the $t$ and $\phi$ components of the geodesics equation (\ref{geodesic}) are integrable and one obtains
\bqn
\tilde E = -g_{\mu\nu} u^{\mu} \xi_{(t)}^{\nu} = f(r) \dot t, \\
\tilde l = g_{\mu\nu} u^{\mu} \xi_{(\phi)}^{\nu} = r^2 \dot \phi,
\eqn
where a dot denotes the derivative with respect to the proper time $\tau$ and $u^{\mu} = \frac{dx^\mu}{d\tau}$ is the four velocity of the particle moving on the geodesic. We have $g_{\mu\nu} u^\mu u^\nu=-\varepsilon$ with $\varepsilon=1$ for massive particle and $\varepsilon=0$ for massless one. For the spherical and static spacetime, the $r$ component of the geodesics equation (\ref{geodesic}) is also integrable. This leads to the equation of motion for $\dot r$ in the form of
\bqn
\dot r^2= g(r) \left( \varepsilon - \frac{\tilde l}{r^2} + \frac{\tilde E^2}{f(r)}\right).
\eqn
Note that for massive particle, $\varepsilon=1$.} Then the geodesic equations can be casted into the following simple forms
\bqn
\dot t &=& - \frac{ \tilde{E}  }{ g_{tt} } = \frac{\tilde{E}}{f(r)}, \lb{tdot}\\
\dot \phi &=& \frac{  \tilde{l}}{g_{\phi\phi}} = \frac{\tilde{l}}{r^2}, \lb{phidot}\\
\ddot r &=& \frac{1}{2}g(r) \Big[-f'(r) \dot t^2+ g^{-2}(r) g'(r) \dot r^2 + 2 r \dot \phi^2\Big].\nb\\
\lb{rddot}
\eqn
\red{Note that }$\tilde E$ is dimensionless and $\tilde l$ has dimension of mass. 

\red{By} integrating the above equations numerically with given initial conditions for the coordinate functions \red{$\{t(\tau_0), r(\tau_0), \phi(\tau_0)\}$ and their derivatives $\{\dot t(\tau_0), \dot r(\tau_0), \dot \phi(\tau_0)\}$,} one can get the orbits of the massive particles in the self-dual spacetime. For the S0-2 star in the galactic center,  its motion can be well approximated by a processing elliptical orbit. Its pericenter advance per orbit due to the relativistic and LQG effects can be computed from Eqs.~(\ref{tdot}, \ref{phidot}, \ref{rddot}), which lead to
 \bqn
 \Delta \phi &\simeq& \frac{6\pi GM}{a (1-e^2)} \left(1- \frac{4}{3}P\right) \nb\\
 &\simeq& \frac{6\pi G_N M}{a (1-e^2)} \left(1 + \frac{8}{3}P\right), \lb{DelP}
 \eqn
 where $a$ is the semimajor axis of the elliptical orbit, $e$ is the eccentricity. Note that by taking $P=0$, one recovers the classical result for the Schwarzschild spacetime. 
 
\section{Data and Data Analysis of the S0-2 star}
\renewcommand{\theequation}{3.\arabic{equation}} \setcounter{equation}{0}

The S0-2 star is a B-type star in the nuclear cluster orbiting the radio source Sgr A* in the galactic center of our galaxy. Its orbit is characterized by an orbital period of 16 years, a semi-major axis of 970 AU and an high eccentricity of  0.88 \cite{GRAVITY1, GRAVITY1a, DoSci}. Over the last three decades the GRAVITY Collaboration have constantly monitored the motion of the S0-2 star and obtained its precise astrometric and spectroscopic data. Recently, they also reported a precision probe for S0-2's orbit including the gravitational redshift and the pericenter advance $\delta \phi \simeq 12 '$ per orbital period. All these data and results are fully consistent with GR. On the other hand, these precise measurements also open a new route to constrain small \red{deviation} from GR in the vicinity of the supermassive black hole.

Here our purpose is to use the publicly available data for the S0-2 star to constrain the self-dual spacetime in LQG. Such constraint can be directly transform to the constraint on the parameters arsing from the LQG itself. 

\subsection{Datasets used in the analysis}

\begin{figure}
\center{
\includegraphics[width=8.1cm]{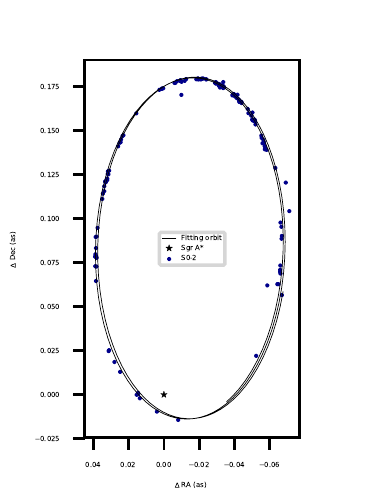}
\caption{The 145 astrometric positions of the S0-2 star (the blue points) and the fitting orbit (the solid line). Readers should be noticed that this orbit is not the real orbit, which is the projection on the sky plane. The unit as is arcseconds and the origin of coordinates is the position of Sgr A*.}
 }  \label{orbit}
\end{figure}

\begin{figure} 
\center{
\includegraphics[width=9cm]{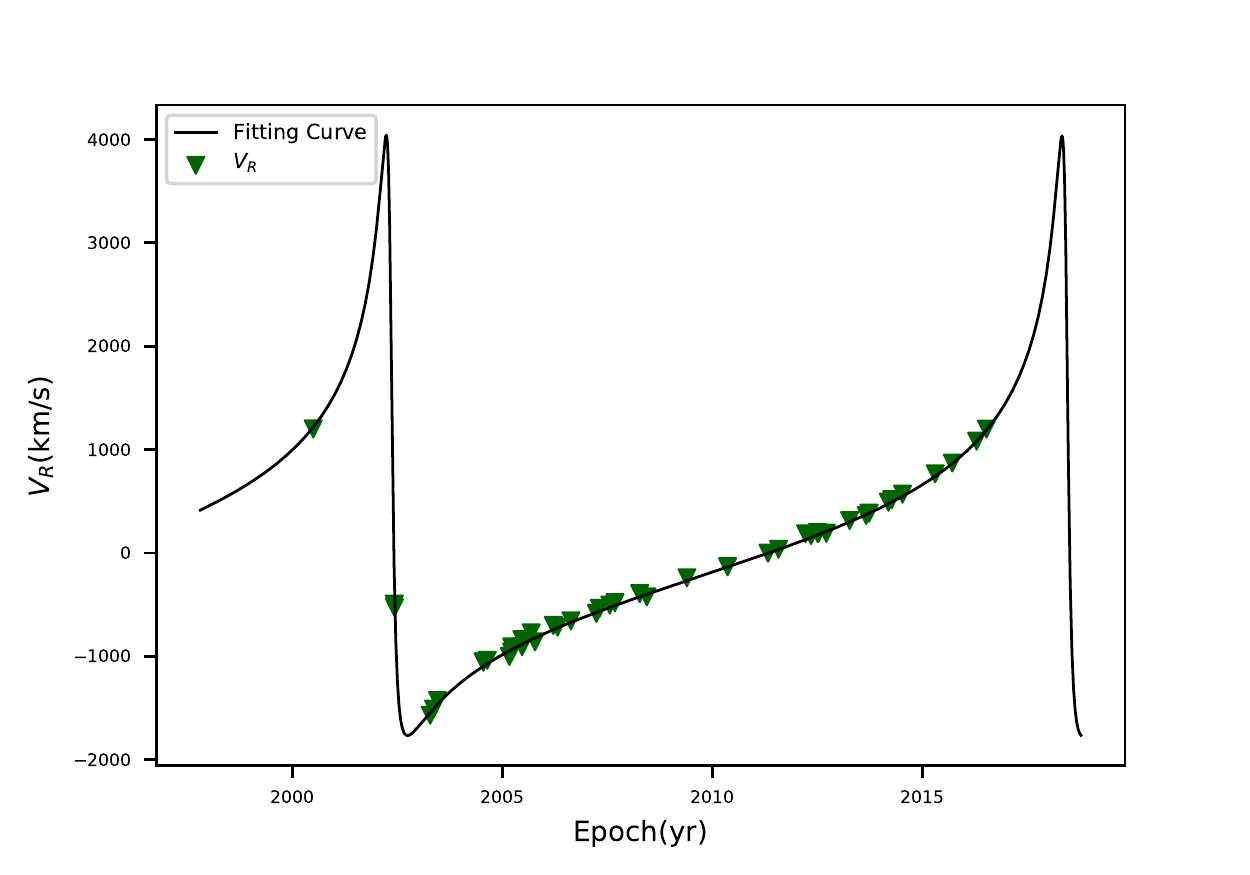}
\caption{The data of radial velocities of the S0-2 star (the green points) and the fitting curve to describe the change of $V_R$ (the solid line).}
 }\label{RV}
\end{figure}

We use the publicly available astrometric and spectroscopic data for the S0-2 star which have been collected over the past decades. These data includes three different parts: the data of astrometric positions, the radial velocities, and the pericenter \red{precession}. The details of these three parts are summarized below: 
\begin{itemize}
    \item[1.] {\em Astrometric positions }: We use 145 astrometric positions of S0-2  between 1992.224 and 2016.53. These data are extracted from \cite{Monitoring}, and the data before 2002 are collected from the ESO New Technology Telescope (NTT) and the others (after 2002) are collected from the Very Large Telescope (VLT). These data are shown in Fig.~\ref{orbit}.
It is worth noting that hereafter the epoch is expressed in Julian year.
\item[2.] {\em Radial velocities}: We use  data of 44 radial velocities between 2000.487 and 2016.519  as reported in   \cite{Monitoring}. These data are also from different observing groups. The data before 2003 are collected from NIRC2 and the others (after 2003) are collected from  the INtegral Field Observations in the Near Infrared (SINFONI). These data are shown in Fig.~\ref{RV}.
\item[3.] {\em Orbital} precession of S0-2: Recently, the GRAVITY Collaboration has measured the orbital precession of S0-2 per orbit \cite{GRAVITY2}
\bqn
\Delta \phi _{per \;orbit}=1.10 \pm 0.19. \lb{OP_12}
\eqn
\end{itemize}
The orbital precession is an important phenomenon predicted by  GR, as we can see it from Fig.\ref{orbit}. In this paper, we  use this measurement in our MCMC analysis to break the degeneracy among some parameters.

\subsection{Modeling the orbit with relativistic effects}

\begin{figure}[h] 
    \centering
    \includegraphics[width=8.1cm]{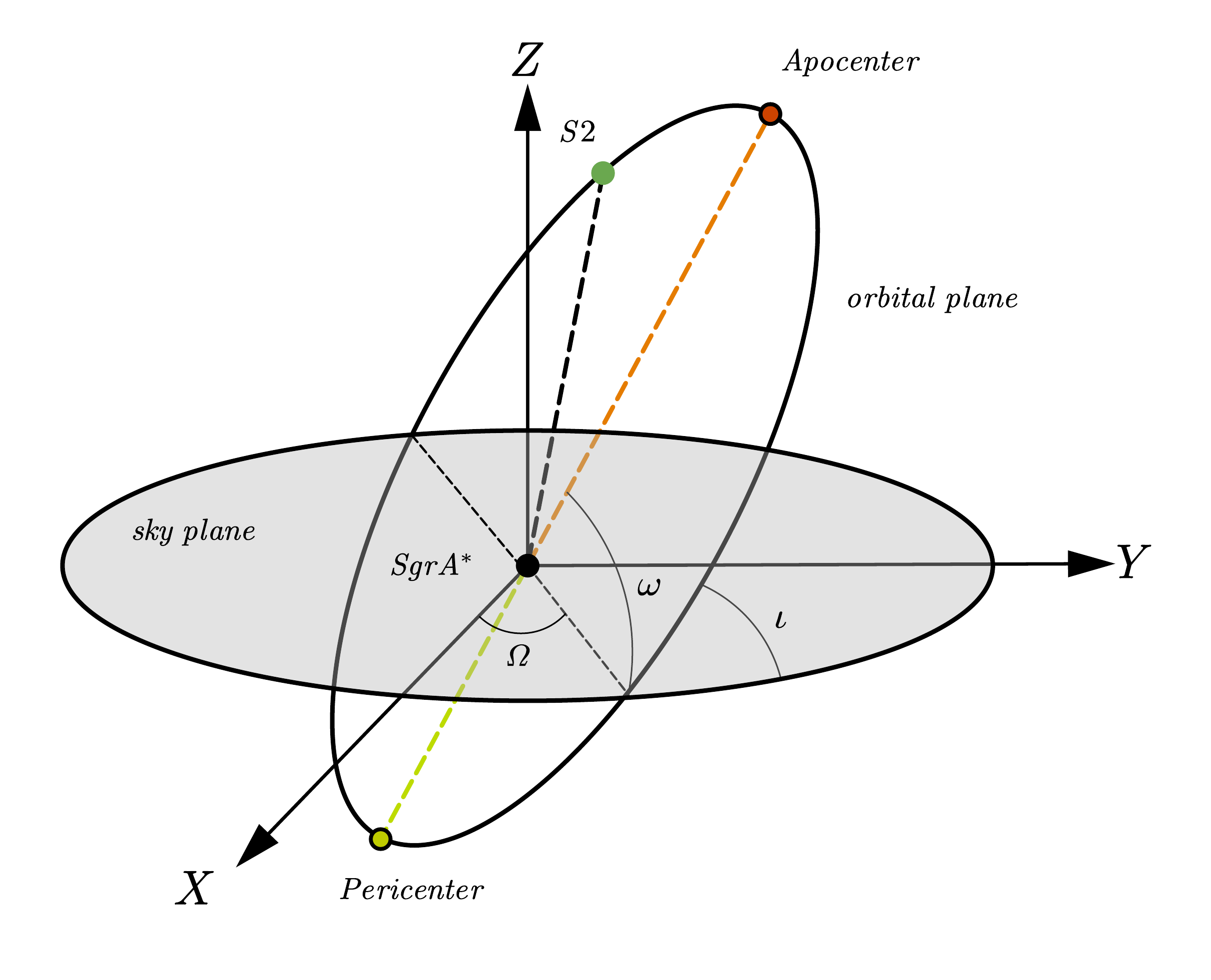} 
    \caption{The astrometric positions lie on the sky plane. The Z-axis follows the direction the Solar system points to the Galactic Center and  the X-axis points East and the Y-axis points North. $\omega$ is the perihelion argument, $\Omega$ the longitude of ascending node, and $\iota$ the orbital inclination  of the processing elliptical orbit for S0-2.}
    \label{cor}
\end{figure}

By integrating Eqs.~(\ref{tdot}, \ref{phidot}, \ref{rddot}) numerically with given initial conditions for coordinates \red{$\{t(\tau_0), r(\tau_0), \phi(\tau_0)\}$ and their derivatives $\{\dot t(\tau_0), \dot r(\lambda_0), \dot \phi(\lambda_0)\}$}, one can get the theoretical positions of S0-2 star in the self-dual spacetime on the orbital plane. However, the astrometric positions of S0-2 we mentioned in the above subsection is described on the sky plane. The relation between the sky plane and the orbital plane is shown in Fig.~\ref{cor}. Here we would like to mention that, in both planes, the motion of S0-2 star can be well approximated by a \red{precessing} elliptical orbit. To compare the theoretical positions and the astrometric positions, we have to make sure all the observational quantities are in the same plane. This can be achieved by projecting the theoretical positions onto the sky plane via
\bqn 
    X &=& x B+y G\lb{Xobs},\\
    Y&=&xA+yF\lb{Yobs},\\
    Z&=&xC+yH \lb{Zobs},
\eqn
where $(X, Y, Z)$ are the coordinates on the sky plane and $(x, y, z)$ are the coordinates on the orbital plane. The coefficients $A, B, C, F, G, H $ can be got from the following formulas
\bqn
    A&=&\cos \Omega \cos \omega -\sin \Omega \sin \omega \cos \iota, \lb{A}\\
    B&=&\sin \Omega \cos \omega +\cos \Omega \sin\omega \cos \iota, \lb{B}\\
    C&=&\sin \omega \sin \iota,  \lb{C}\\
    F&=&-\cos\Omega \sin \omega -\sin \Omega \cos \omega \cos \iota, \lb{F}\\
    G&=&-\sin \Omega \sin \omega +\cos \Omega \cos \omega \cos \iota,  \lb{G}\\
    H&=&\cos \omega \sin \iota, \lb{H}
\eqn
where $\omega$ is the perihelion argument, $\Omega$ the longitude of ascending node, and $\iota$ the orbital inclination  of the processing elliptical orbit for S0-2. In Fig.~\ref{cor}, we illustrate the relation between the sky and orbital plane and the geometric meanings of $\Omega$, $\omega$, and $\iota$.

Furthermore, considering there are  offsets and  linear drifts between the gravitational center and the reference frame,  we need to introduce  $x_{0},y_{0},v_{x0},v_{y0}$  to model it ~\cite{DoSci}
\bqn
    X&=&X(t_{\rm em})+x_{0}+v_{\rm x0}(t_{\rm em})(t_{\rm em}-t_{\rm refer}),  \lb{x0} \\  
    Y&=&Y(t_{\rm em})+y_{0}+v_{\rm y0}(t_{\rm em})(t_{\rm em}-t_{\rm refer}),  \lb{y0}
\eqn
where $t_{\rm refer}$ is the reference epoch for the parameters $x_{0},y_{0},v_{x0},v_{y0}$ and $t_{\rm em}$ is the epoch when the light emit. Here $x_{0},y_{0}$ mean the offsets and  $v_{\rm x0}(t_{\rm em})(t_{\rm em}-t_{\rm refer}), v_{\rm y0}(t_{\rm em})(t_{\rm em}-t_{\rm refer})$ mean the linear drifts. \red{For our data, the reference year $t_{\rm refer}=2009.2$ \cite{Pinpointing}}.

We also need to consider several relativistic effects for comparing the theoretical positions with astrometric data.

We first consider the \red{effects} of the Romer time delay, which modulates the time of arrival of the light emitted by the star when is farther away or closer to Earth during its orbital motion. The Romer time delay can be expressed as
\bqn
t_{\rm obs}-t_{\rm em}=\frac{Z(t_{\rm em})}{c},
\eqn
where $t_{\rm obs}$ is the epoch when we observe the light and $Z$ can be obtained from Eqs.(\ref{Zobs}). This equation is difficult to solve so one can use an iteration scheme to solve the equation \cite{DoSci, GRAVITY1}:
\bqn
    t_{\rm em}^{(i+1)}=t_{\rm obs}-\frac{Z(t_{\rm em}^{(i)})}{c}.
\eqn
For our purpose, after one iteration this equation becomes,
\bqn
    t_{\rm em} \approx t_{\rm obs}-\frac{Z(t_{\rm obs})}{c}.
\eqn

Then let us turn to  the effects of photon's frequency shift $\zeta$ which can be related to the radial velocity of the S0-2 star
\bqn
    \zeta = \frac{\Delta \nu}{\nu} = \frac{\nu_{\rm em}-\nu_{\rm obs}}{\nu_{\rm obs}}=\frac{V_{\rm R}}{c},
\eqn
where $\nu_{\rm em}$ is the frequency of light when it emits, $\nu_{\rm obs}$ is the frequency when one observes it, and $V_{\rm R}$ is the radial velocity of the S0-2 star.  Two relativistic effects can make important contributions to the above frequency shift. One is the Dopple shift $\zeta_{\rm D}$ and the other is the gravitational redshift $\zeta_G$.

We first consider the Doppler shift $\zeta_{\rm D}$ which is caused by relative motion between  the star and the observer. Due to the high velocity of  S0-2, the Doppler shift would cause great impact here 
\bqn
    \zeta_{\rm D}=\frac{\sqrt{1-\frac{v_{\rm em}^{2}}{c^{2}}}}{1-\bm n \cdot \bm v_{\rm em}},
\eqn
where $v_{em}$ is the velocity  at $t_{em}$,  $\bm n \cdot \bm v_{\rm em}$ is the velocity projected on to the sight of the light, which is  known as the radial velocity. The  gravitational shift $\zeta_{\rm G}$ is a GR effect and  this frequency shift would become significant under strong gravitational fields which reads
\bqn
\zeta_{\rm G}=\frac{1}{\sqrt{-g_{00}}}.
\eqn
 Therefore, we get 
\bqn
\zeta= \zeta_{\rm D} \cdot \zeta_{\rm G}-1.
\eqn

Also, considering that the Sgr A* might moving toward the sun, which would affect the velocity, we introduce $v_{\rm z0}$ here to model the $V_{\rm R}$ \cite{Reid:2006cu}.
\bqn
V_{\rm R}=c \cdot \zeta + v_{\rm z0}.
\eqn

\subsection{Analysis of Monte Carlo Markov Chain}

In this paper, we carry out the analysis of the MCMC implemented by \textit{emcee} \cite{emcee} to obtain the constraints on the self-dual spacetime. We explore the following parameters
\bqn
\{M, R_{0}, a, e, \iota, \omega, \Omega, t_{\rm apo}, x_{0}, y_{0}, v_{x_{0}}, v_{y_{0}}, v_{z_{0}}, P\},\lb{paras} \nb\\
\eqn
to fit the theoretical orbits to the publicly available data as described in Sec.III.A. Here
$M$ and $R_0$ are the mass of the central black hole and the distance between the Earth and the black hole, $\{ a, e, \iota, \omega, \Omega, t_{\rm apo}\}$ are the six orbital elements which describe the osculating elliptical orbit of the S0-2 star. The five parameters $\{x_{0}, y_{0}, v_{x_{0}}, v_{y_{0}}, v_{z_{0}}\}$ represent the zero-point offsets and drifts of the reference frame with respect to the mass centroid, and $P$ is the polymeric function arsing from the self-dual spacetime in LQG. Here we would like to note that the orbit of S0-2 is not  an exact ellipse, but a processing ellipse. For every single point of the orbit  one can associate  a corresponding ellipse which is called the osculating ellipse and described by the above orbital elements. 

To carry out our MCMC analysis with the above parameter space, \red{we use two prior sets, the Gaussian and uniform priors respectively, for the 13 orbital parameters. The Gaussian priors are set to be centered on the best values given in \cite{GRAVITY2}. For the polymeric function $P$ in the self-dual spacetime, we use uniform prior for both simulations.}
The prior sets used for our MCMC analysis is summarized in Table.~\ref{priors}. Similar prior sets have also been used in constraining different modified theories of gravity with astrometric data of S0-2  \cite{Borka:2021omc, deMartino:2021daj, DellaMonica:2021xcf, DellaMonica:2021fdr, DAddio:2021smm}.

\begin{table}
\caption{\label{priors}
\red{Two different prior sets used for our MCMC analysis. The Gaussian priors are set to be centered on the best values given in \cite{GRAVITY2}. In this table, the units $M_{\rm \odot}$ is the solar mass, kpc is the kiloparsec, mas is the milliarcsecond, $^{\circ}$ is the degree, and yr is the year.} }
\begin{ruledtabular}
\begin{tabular}{lccc}
&\multicolumn{2}{c}{Gaussian prior} &\multicolumn{1}{c}{Uniform prior}\\
\cline{2-3}  \cline{4-4}
 Paramaters& $\mu$ & $\sigma$ & - \\
  \colrule
     $ M \;(10^{6}M_{\rm \odot})$ &    4.261 & 0.012  & [0, 10] \\
     $R_{0}$ (kpc) & 8.2467  & 0.0093 & [5, 10] \\
     $a$ (mas) & 125.058    & 0.04 & [120, 130]\\
     $e$ & 0.884649   & 0.00008 & [0.5, 1.5] \\
     $\iota$ ($^{\circ}$)  & 134.567    & 0.033  & [130, 140] \\
     $\omega$ ($^{\circ}$) & 66.263 & 0.031 &  [60, 70] \\
     $\Omega$ ($^{\circ}$)  & 228.171 & 0.031 & [220, 240] \\
     $t_{\rm apo}$ (yr) &  2010.38  & 0.00016 & [2009, 2011]\\
     $x_{0}$ (mas) & -0.90   & 0.14  & [-50, 50] \\
     $y_{0}$ (mas) & 0.07   & 0.12  & [-50, 50]\\
     $v_{\rm x_{0}}$ (mas/yr) & 0.080   & 0.010 & [-50, 50]\\
     $v_{\rm y_{0}}$ (mas/yr) & 0.0341  & 0.00096 & [-50, 50]\\
     $v_{\rm z_{0}}$ (km/s) & -1.6 & 1.4 & [-50, 50] \\
     \colrule
     $P$ & - & - &[0, 1] \\
\end{tabular}
\end{ruledtabular}
\end{table}

As we mentioned in Sec.III.A, three different parts of data are employed in our MCMC analysis. For this reason, the likelihood function $\mathcal{L}$ consists of three parts, i.e.
\begin{eqnarray}
\log {\cal L} = \log {\cal L}_{\rm AP} + \log {\cal L}_{\rm VR} + \log {\cal L}_{\rm pro},\lb{likelyhood}
\end{eqnarray}
where $\log {\cal L}_{\rm AP}$ denotes the likelihood of the 145 astrometric positional data
\bqn
\log {\cal L}_{\rm AP} &=& - \frac{1}{2} \sum_{i} \frac{(X_{\rm obs}^i -X_{\rm the}^i)^2}{(\sigma^i_{X,{\rm obs}})^2} \nb\\
&& -\frac{1}{2} \sum_{i} \frac{(Y_{\rm obs}^i -Y_{\rm the}^i)^2}{(\sigma^i_{Y,{\rm obs}})^2},
\eqn
$\log {L}_{\rm VR}$ represents the likelihood of the 45 data of the radial velocities
\begin{eqnarray}
\log {L}_{\rm VR} - \frac{1}{2} \sum_{i} \frac{(V_{\rm R, obs}^i - V_{\rm R, the}^i)^2}{(\sigma^i_{V_{\rm R, obs}})^2},
\end{eqnarray}
and $\log {\cal L}_{\rm pro}$ is the likelihood of the orbital \red{precession}
\bqn
\log {\cal L}_{\rm pro} = - \frac{1}{2} \frac{(\Delta \phi_{\rm obs}-\Delta \phi_{\rm the})^2}{\sigma^2_{\Delta \phi, {\rm obs}}}.
\eqn
Here $X_{\rm obs}^i$, $Y_{\rm obs}^i$, and $V_{\rm R, obs}^i$ are the data of the astrometric positions and radial velocities, and $X_{\rm the}^i$, $Y_{\rm the}^i$, and $V_{\rm R, the}^i$ are the corresponding theoretical predictions. $\Delta \phi_{\rm obs}$ represents the measurements of the orbital \red{precession} given in (\ref{OP_12}) and $\Delta \phi_{\rm the}$ is given by (\ref{DelP}) with given values of $P$. In the above expressions, $\sigma^i_{x, {\rm obs}^i}$ denote the corresponding statistical uncertainty for the associated quantities.

\subsection{Results and Discussions}

\begin{figure*}
    \centering
    \includegraphics[width=17.2cm]{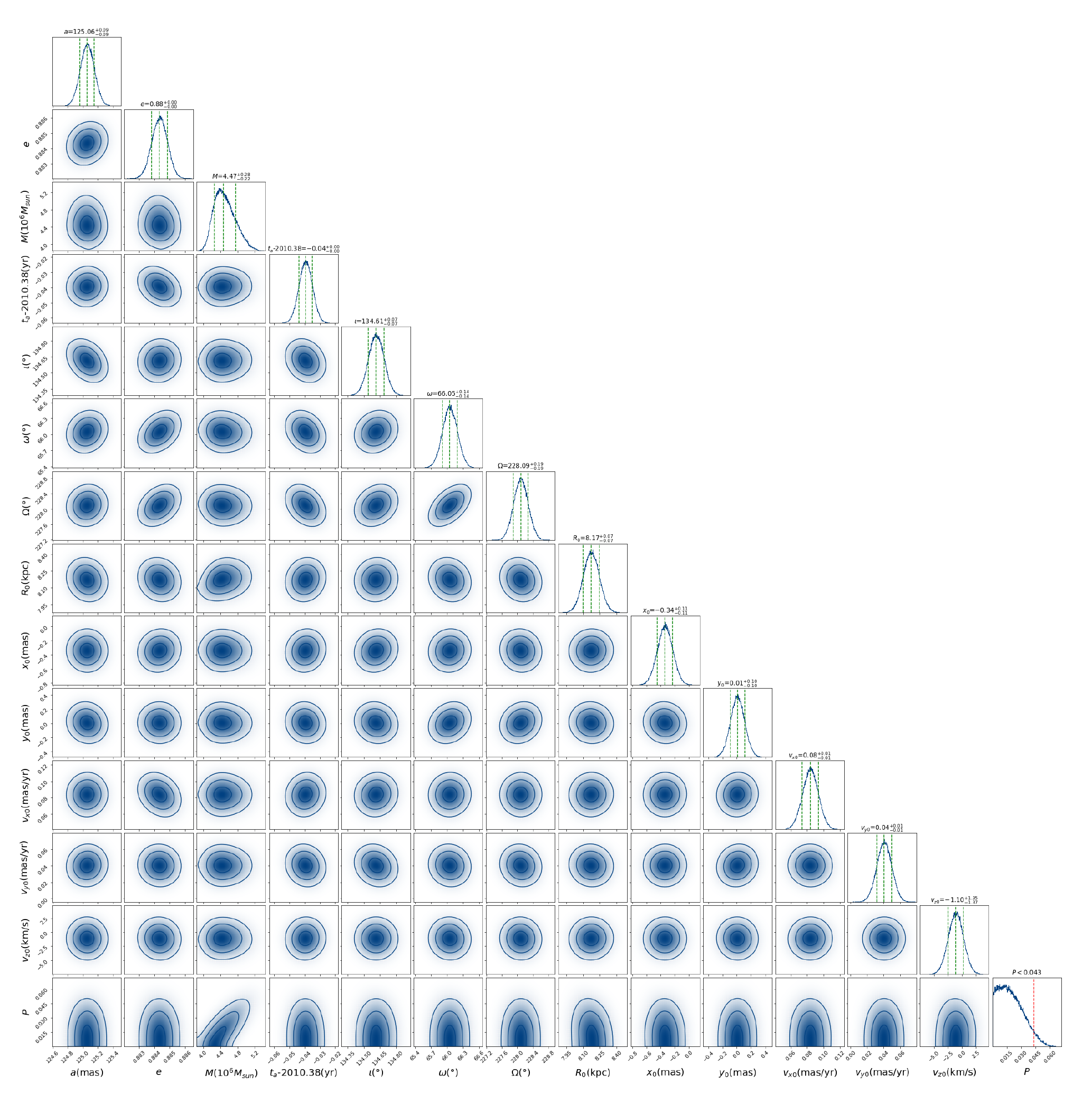}
    \caption{\red{The posterior distribution of the orbital parameters of the S0-2 star and the polymeric function $P$ of the self-dual spacetime with Gaussian priors for the orbital parameters.}}
    \label{contour1}
\end{figure*}

\begin{figure*}
    \includegraphics[width=17.2cm]{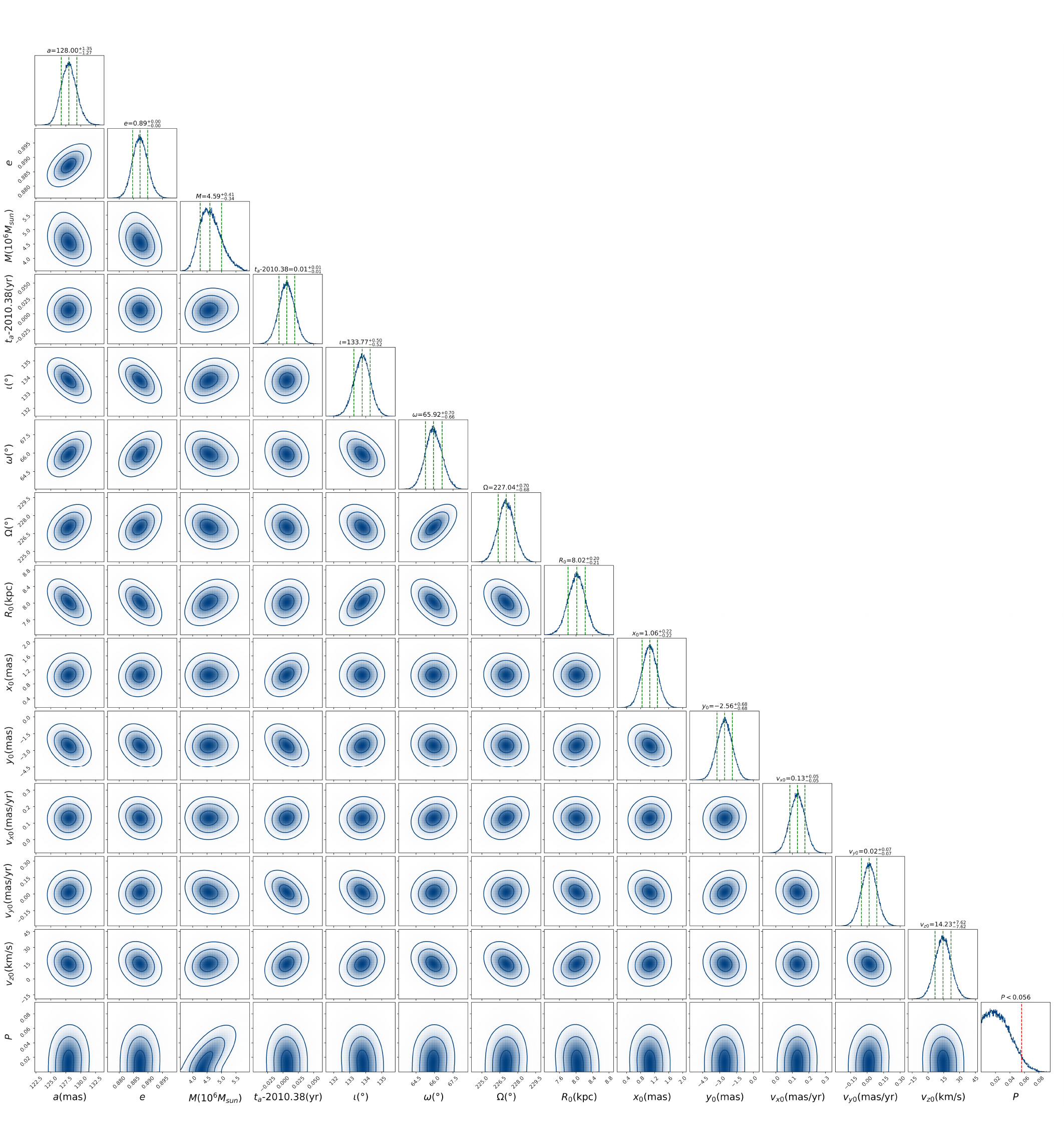}
    \caption{\red{The posterior distribution of the orbital parameters of the S0-2 star and the polymeric function $P$ of the self-dual spacetime with uniform priors for the orbital parameters.}}
    \label{contour2}
\end{figure*}

With the setup described in the above subsections, we explore the above mentioned 14-dimensional parameter space through an analysis of MCMC. In Fig.~\ref{contour1} and \red{Fig.~\ref{contour2},} we illustrate the full posterior distributions of our 14-dimensional parameter space of our orbital model for two sets of priors respectively. On the contour plots of this figure, the shaded regions show the 68\%, 90\%, and 95\% confidence levels (C.L.) of the posterior probability density distributions of the entire set of parameters, respectively. The corresponding best fit values of these 14 parameters for two sets of priors are presented in Table.~\ref{para}. The comparison of the orbit of these best-fit values and the astrometric data is also presented in Fig.~\ref{orbit}. The results are presented in \red{Fig.~\ref{contour1} ,  Fig.~\ref{contour2}} and Table.~\ref{para} .


\begin{table}
\caption{\label{para}
The best-fit values of the parameters of the orbital model of S0-2 in the self-dual spacetime, resulting from the MCMC analysis. The upper bound on $P$ is derived from the posterior region at 95\% C.L..}
\begin{ruledtabular}
\begin{tabular}{lcc}
Parameters &\multicolumn{2}{c}{Best-fit values}  \\
\cline{2-3}
           & Gaussian & uniform\\
\colrule
     $ M\; (10^{6}M_{\odot})$ &    4.38 & 4.46 \\
     $R_{0}$ (kpc) & 8.17  &8.02 \\
     $a$ (mas) & 125.06 &127.98\\
     $e$ & 0.8844  &0.8872\\
     $\iota$ ($^{\circ}$) & 134.61 &133.72\\
     $\omega$ ($^{\circ}$)& 66.03 &65.84\\
     $\Omega$ ($^{\circ}$) & 228.102 &226.97 \\
     $t_{\rm apo}$ (yr) &  2010.34 &2010.38\\
     $x_{0}$ (mas) & -0.35 &1.06 \\
     $y_{0}$ (mas) & 0.0005 &-2.51\\
     $v_{\rm x_{0}}$ (mas/yr) & 0.084 & 0.134\\
     $v_{\rm y_{0}}$ (mas/yr) & 0.041 & 0.020\\
     $v_{\rm z_{0}}$ (km/s) & -0.96 & 13.42\\
     \colrule
     $P$   &   $<$ 0.043 & $<$ 0.056
\end{tabular}
\end{ruledtabular}
\end{table}

\begin{figure}
    \centering
    \includegraphics[width=8.1cm]{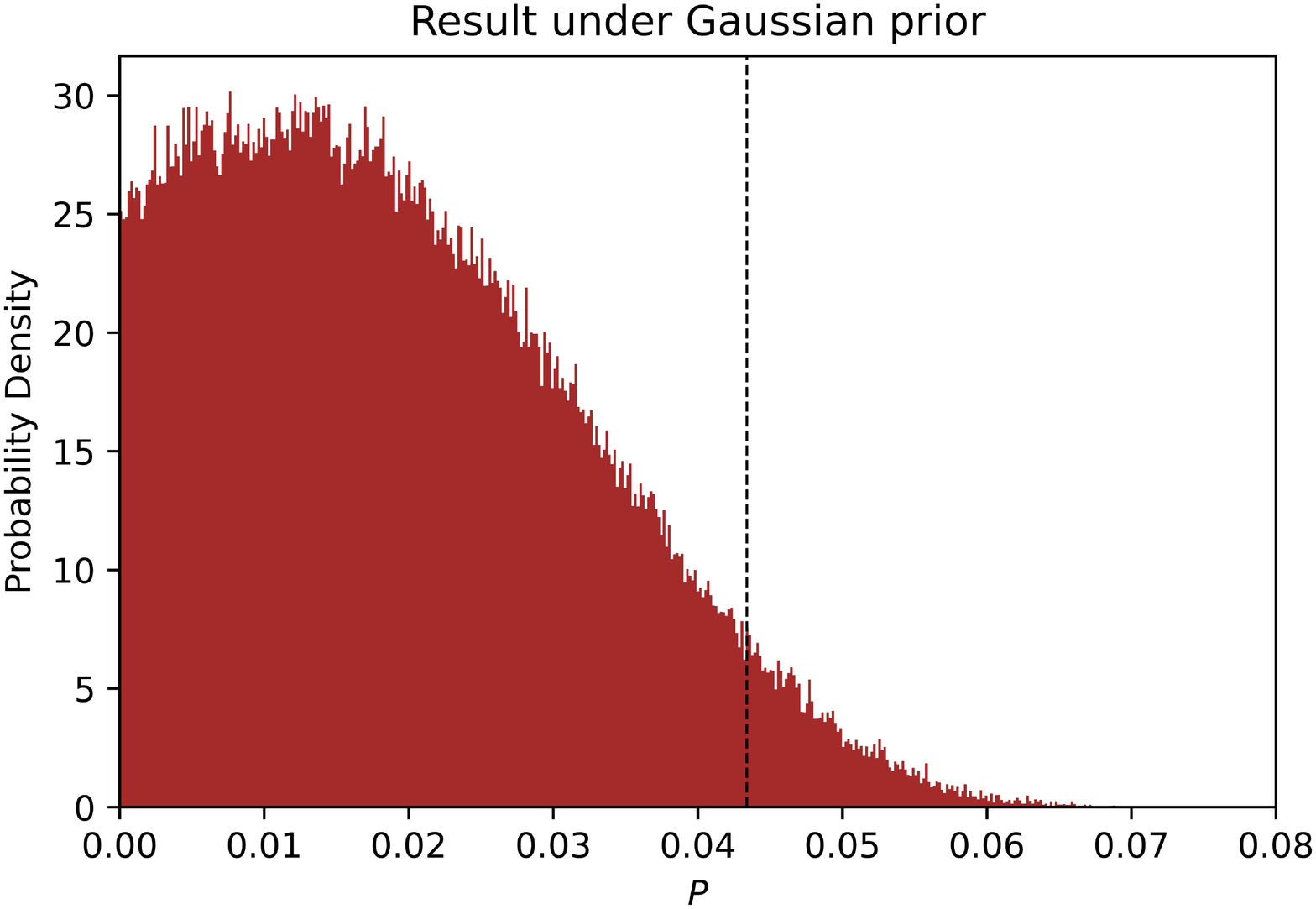}
    \caption{This figure is picked from Fig.~\ref{contour1} to show more details. The  upper limit of $P$ is 0.043 at 95\% confidence level.}
    \label{P1}
\end{figure}

\begin{figure}
    \centering
    \includegraphics[width=8.1cm]{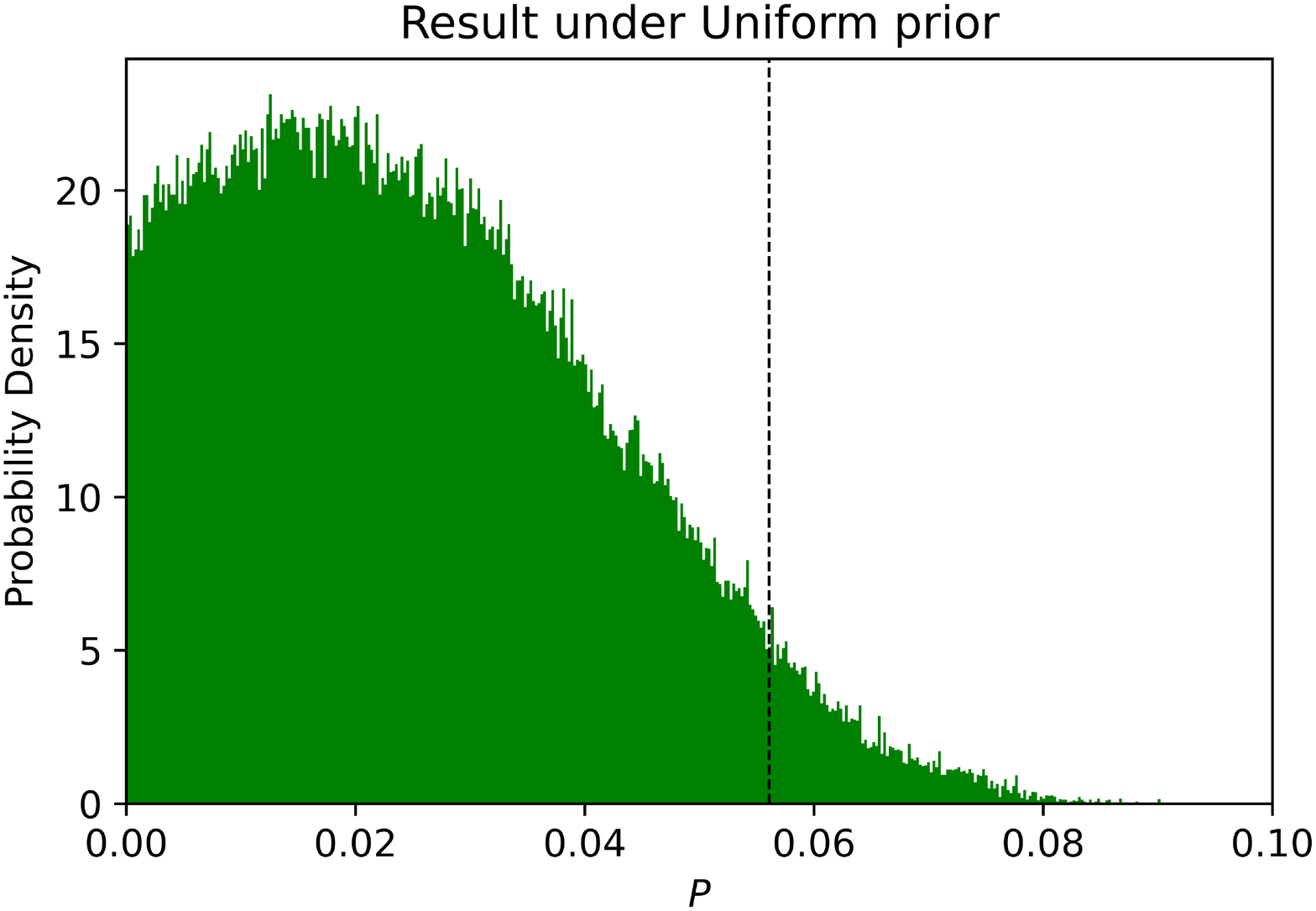}
    \caption{This figure is picked from Fig.~\ref{contour2} to show more details. The  upper limit of $P$ is 0.056 at 95\% confidence level.}
    \label{P2}
\end{figure}

In order to estimate the observational constraint on the polymeric function $P$ of LQG, we plot the marginalized posterior distribution of $P$ in Fig.~\ref{P1} and Fig ~\ref{P2}. Then the upper bounds on $P$ can be calculated from the corresponding posterior distribution of $P$. We find the polymeric function $P$ can be constrained at 95\% confidence level to be
\begin{eqnarray}
P < 0.043,
\end{eqnarray}
for Gaussian prior and
\begin{eqnarray}
P < 0.056,
\end{eqnarray}
for uniform prior respectively. 

The \red{constraints} we obtain here \red{are both} a little bit stronger (but compatible) than the bound $P<0.067$ obtained in \cite{Zhu:2020tcf} by directly using the error of the measurement of the orbital \red{precession} of the S0-2 star. However, it is much weaker than those obtained from the measurement of the gravitational time delay by the Cassini mission, the deflection angle of light by the Sun, and the perihelion advance of Mercury \cite{Zhu:2020tcf}. Although the observational constraints from the S0-2 is not accurate enough with respect to other type observations, our results show that the observational data from the observations of S0-2 star at the galactic center does have the capacity for constraining black hole parameters beyond those in GR. In addition, we would like to mention that those tighter constraints mentioned above are all derived from the observations at the scale of the solar system, our results represent a bound on $P$ from a very different environment of strong gravity regime.

\section{Conclusions}
\renewcommand{\theequation}{5.\arabic{equation}} \setcounter{equation}{0}

In this paper, we consider the effects of the self-dual spacetime of LQG on the orbit of the  S0-2 star orbiting Sgr A* in the central region of our Milky Way. The effects of LQG may not only lead to signatures on the orbits of the S0-2 star, but also affect its overall peri- centre advance. We also compare the effects of the self-dual spacetime with the publicly available astrometric and spectroscopic data, including the astrometric positions, the radial velocities, and the orbital \red{precession} for the S0-2 star. With these data, we perform a \red{MCMC} simulation to probe the possible LQG effects on the orbit of the S0-2 star. We do not find any significant evidence of the self-dual space-time and thus place an upper bound at 95\% confidence level on the polymeric function $P < 0.043$ and $P<0.056$, corresponding to the Gaussian and uniform priors for orbital parameters, respectively. This bounds lead to a constraints on the polymeric parameter $\delta$ of LQG to be $| \delta| < 1.82$ and $|\delta|<2.11$ respectively.  Finally, we would like to mention that we only consider the static self-dual spacetime in this paper and ignore the effects of the angular momentum of the spacetime. For all the observational effects we considered in this paper, the effects due to rotation of the central black hole are expected to be very small.


\section*{Acknowledgements}

This work is supported by the Zhejiang Provincial Natural Science Foundation of China under Grants No. LR21A050001 and No. LY20A050002, the National Natural Science Foundation of China under Grants No. 11675143 and No. 11975203, the National Key Research and Development Program of China under Grant No. 2020YFC2201503, and the Fundamental Research Funds for the Provincial Universities of Zhejiang in China under Grant No. RF-A2019015.

\end{document}